# Scalable colored sub-ambient radiative coolers based on a polymer-Tamm photonic structure

*Tianzhe Huang,[1,3,*] Qixiang Chen,[2,*] Jinhua Huang,[1] Yuehui Lu,[1,†] Hua Xu,[2] Meng Zhao,[4] Yao Xu,[5] and Weijie Song[1,‡]*

[1]Zhejiang Provincial Engineering Research Center of Energy Optoelectronic Materials and Devices, Ningbo Institute of Materials Technology and Engineering, Chinese Academy of Sciences, Ningbo 315201, China

[2]School of Physical Science and Technology, Ningbo University, Ningbo 315211, China

[3]Nano Science and Technology Institute, University of Science and Technology of China, Suzhou 215123, China

[4]Jiangsu Key Laboratory of Micro and Nano Heat Fluid Flow Technology and Energy Application, Suzhou University of Science and Technology, Suzhou 215009, China

[5]Xi'an Institute of Optics and Precision Mechanics, Chinese Academy of Sciences, Xi'an 710119, China

---

[*] T. Huang and Q. Chen contributed equally to this work.

[†] Corresponding author: yhlu@nimte.ac.cn

[‡] Corresponding author: weijiesong@nimte.ac.cn

**SUMMARY**

Daytime radiative coolers cool objects below the air temperature without any electricity input, while most of them are limited by a silvery or whitish appearance. Colored daytime radiative coolers (CDRCs) with diverse colors, scalable manufacture, and sub-ambient cooling have not been achieved. We introduce a polymer-Tamm photonic structure to enable a high infrared emittance and an engineered absorbed solar irradiance, governed by the quality factor (Q-factor). We theoretically determine the theoretical thresholds for sub-ambient cooling through yellow, magenta, and cyan CDRCs. We experimentally fabricate and observe a temperature drop of 2.6-8.8 °C on average during daytime and 4.0-4.4 °C during nighttime. Furthermore, we demonstrate a scalable-manufactured magenta CDRC with a width of 60 cm and a length of 500 cm by a roll-to-roll deposition technique. This work provides guidelines for large-scale CDRCs and offers unprecedented opportunities for potential applications with energy-saving, aesthetic, and visual comfort demands.

**INTRODUCTION**

Global climate change and energy consumption are two great challenges that human society faces in the 21st century.[1] Most electricity is generated by fossil fuel and used for space cooling, which is accompanied by excessive greenhouse gas emission and temperature rise.[2] Air conditioners account for 15% of global electricity consumption and 10% of greenhouse gas emission.[3] An eco-friendly and energy-effective cooling strategy is appealing from a green and sustainable development outlook. Daytime radiative cooler (DRC), firstly proposed by Raman,[4] consumes no electricity to reduce the temperature of objects below that of ambient air, the so-called sub-ambient cooling, by strongly reflecting the solar irradiance at $\lambda = 0.3\text{-}4$ μm and simultaneously radiating heat out to the extremely cold universe through an atmospheric transparency window at $\lambda = 8\text{-}13$ μm.[5] A variety of DRCs have been demonstrated, which are promising for potential applications in energy-efficient buildings,[6] photovoltaic modules,[7] and personal thermal management.[8] However, the majority of the proposed DRC materials exhibit white or metallic colors for guaranteeing a strong reflection of the solar irradiance,[9] which is unfavorable to some application scenarios and installation locations such as building envelopes[10] and vehicles[11] with aesthetic purposes and potentially leads to light pollution and eye safety issues.

The realization of colored daytime radiative coolers (CDRCs) with diverse colors is appealing but highly challenging. Though some theoretical proposals have been raised,[12-15] experimental demonstration is still few. The recent efforts to experimentally realize CDRCs mainly adopted the strategies of dye coloring, structural color, and photoluminescence (PL).[16-20] Dyes and pigments enable the diverse colors, while are not good candidates for sub-ambient cooling due to their considerable absorption in the near-to-short wavelength infrared at 0.74-2.5 μm.[16] Even though a bilayer structure, composed of a visible-absorptive pigment layer atop a common solar-scattering layer, is employed, the bilayer coatings just stay cooler than the commercial paint monolayers of the same color and still warmer than the air ambient.[17] Alternatively, structural color enables diverse colors based on optical interference and diffraction with the solar non-absorptive materials. Lee et al. fabricated a structural colored structure based on metal-insulator-metal (MIM) layers,[18] which could not achieve sub-ambient cooling unless a spatial pattern was made, equivalently reducing the colored area and mitigating the thermal load due to solar absorption. To prevent CDRCs from being considerably heated by the absorbed solar irradiance, Son et al. employed the silica-embedded perovskite nanocrystals to excite PL effect, successfully cooling to 1.7-4.2 °C below the ambient air temperature.[19] Yoon et al. adopted the Cu-based QDs to compose the CDRCs and achieved a sub-ambient temperature drop of 0.51-3.25 °C.[20] Despite the enabled sub-ambient cooling by the PL-based CDRCs, they suffer from the limited colors and scalable manufacture as well as potential durability issues. Although these efforts significantly improve CDRCs, the theoretical insight to what extent a specific color determines sub-ambient radiative cooling performance has not been explicitly provided. On the other hand, comprehensively enabling diverse colors, sub-ambient cooling, and scalable manufacture remains unsolved, severely hindering the large-scale deployment of CDRCs for practical applications. In terms of both scientific and application points of view, it is significant to develop scalable-manufactured CDRCs with diverse colors while achieving sub-ambient cooling.

In this work, to enable scalable-manufactured CDRCs with diverse colors for all-day sub-ambient cooling, we introduce a polymer-Tamm photonic structure, in which a high infrared emittance is endowed for thermal radiation and the absorbed solar irradiance and colors can be elaborately engineered by tailoring the spectral width and strength of Tamm plasmon resonances (TPRs), characterized by the quality factor (Q-factor). Staring from an ideal CDRC, we performed a detailed theoretical analysis of a realistic CDRC through modeling a modified state-of-the-art DRC and explicitly determined the Q-factor thresholds of 9.06, 10.07, and 10.86 for sub-ambient yellow, magenta, and cyan CDRCs, respectively. We fabricated the CDRCs with the Q-factors exceeding the theoretical thresholds, not only exhibiting the subtractive primary colors of cyan, magenta, and yellow (CMY) and, but also cooling to 2.6-8.8 °C and 4.0-4.4 °C below the air temperature on average during daytime and nighttime, respectively. We furthermore demonstrated a 60-cm-wide and 500-cm-long large-area magenta CDRC by using a roll-to-roll (R2R) magnetron sputtering deposition and a blade coating for scalable manufacture. This work provides both theoretical and experimental guidelines for achieving the scalable-manufactured CDRCs by handling the dilemmas in color and thermal load while offering unprecedented opportunities for potential applications in building envelopes and vehicles that have energy-saving, aesthetic, and visual comfort demands.

**RESULTS AND DISCUSSION**

**Theoretical analysis**

An ideal DRC possesses a high solar reflectance in the solar spectrum (0.3-2.5 μm) and a high emittance in the long-wave infrared region (> 8 μm), minimizing solar absorption and radiating heat to the cold outer space through the primary atmospheric window at 8-13 μm, respectively, for cooling objects down to sub-ambient temperatures.[21] Thus, the DRCs generally present a silvery or whitish appearance for sufficiently reflecting solar irradiance by introducing metal mirrors or hierarchically porous structures.[22-24]

To achieve the sub-ambient CDRCs, the effects of colorants on the solar reflectance and thermal emissivity must be considered. In contrast to the coloration by chemical dyes or pigments, the structural color can be realized by adopting inherently solar-transparent materials, introducing a less absorbed solar irradiance[25]. For an ideal CDRC with a subtractive primary color of yellow, the solar reflectance spectrum is presented in Figure 1A, together with the infrared absorptance/emittance spectrum. The yellow CDRC has a reflectance dip at 435 nm within the visible light wavelength range (400-700 nm), corresponding to a strong absorption complementary to the color of yellow, while maintaining a high reflectance in the solar spectrum elsewhere. On the other hand, the CDRC possesses either a selective thermal emission only in the atmospheric transparency window or a broadband emission over the MIR spectral range: the former has a unity emittance at 8-13 μm and a zero emittance elsewhere; the latter has a broadband unity emittance throughout the wavelength range from 2.5 to 20 μm. The magenta and the cyan CDRCs have similar spectra expect the different wavelengths of reflectance dips (see Supplementary Figure S1). To balance color display and thermal load, the relationship between the visible light reflection dips or absorption complementary to the desired colors and the net cooling power need to be comprehensively understood. The net cooling power is expressed by[4]

$$P_{\text{cool}} = P_{\text{rad}}(T) - P_{\text{atm}}(T_{\text{amb}}) - P_{\text{sun}}(\theta) - P_{\text{cond+conv}} \quad (1)$$

where $T$ is the surface temperature of the radiative cooler and $T_{\text{amb}}$ is the ambient temperature. $P_{\text{rad}}(T)$ is the radiative power of the cooler, and $P_{\text{atm}}(T_{\text{amb}})$ is the absorbed atmospheric thermal radiation at $T_{\text{amb}}$. $P_{\text{sun}}(\theta)$ is the power of solar irradiation absorbed by the cooler. $P_{\text{cond+conv}}$ is the absorbed power due to convective and conductive heat exchange.

To understand the effects of color display on the absorbed solar irradiance $P_{\text{sun}}(\theta)$, we assume an ideal zero-reflectance dip with a Gaussian line in the visible light range, the bandwidth of which is represented by the quality factor (Q-factor), defined as $\lambda_{\text{max}}/\Delta\lambda$ (i.e., the ratio of the central wavelength of the reflectance dip $\lambda_{\text{max}}$ to the full-width at half-maximum (FWHM) $\Delta\lambda$).[12,26] The central wavelengths for the subtractive primary colors of CMY are in general around 700, 550, and 435 nm, respectively.[12] For the yellow, magenta, and cyan CDRCs, the dependences of the net cooling power $P_{cool}$ on the Q-factor of the reflectance dip and the surface temperatures of radiative coolers $T$ can be determined, as revealed in Figure 1B. The steady-state temperature of radiative coolers $T_s$ is reached when the net cooling power $P_{cool}$ approaches zero. The sub-ambient cooling suggests that the steady-state temperature is lower than the ambient air temperature, i.e., $T_{amb} = 300$ K. The yellow CDRC shows $T_s = T_{amb}$ when $Q = 3.2$, which is the theoretical threshold of Q-factor for sub-ambient cooling, corresponding to an absorbed solar irradiance of $P_{sun} = 173.59$ W/m². As increasing the Q-factor, the absorbed solar irradiance $P_{sun}$ decreases such that the lower steady-state temperatures can be reached. For instance, the steady-state temperature $T_s = 290$ K when $Q = 35$. Similarly, the theoretical thresholds of Q-factor for sub-ambient magenta and cyan CDRCs can be determined to be 5.2 and 5.4, respectively. It reveals that the pivotal challenge for sub-ambient CDRCs is to introduce a reflectance dip with a Q-factor exceeding the theoretical threshold at the desirable wavelength without compromising the reflectance elsewhere. Moreover, in the case of a non-zero reflectance dip in the visible light range, the theoretical thresholds also can be determined (see Supplementary Figure S2). For the selectively emissive CDRCs that has a unity emittance within the atmosphere transparency window from 8 to 13 μm and zero outside, both the theoretical thresholds of Q-factor can be determined following the similar procedures (see Figure S3, Supporting Information). The pathway towards achieving the sub-ambient CDRCs has been elucidated and the effects of the engineered Q-factor on color display, e.g., chroma and hue, are further investigated. It is known that the color hue, e.g., yellow, magenta, and cyan, is governed by the central wavelength of the reflection dip $\lambda_{max}$ and is independent of the Q-factor. Instead, the color chroma is closely related to the Q-factor, as displayed in Figure 1C for the yellow, magenta, and cyan CDRCs. Decreasing the Q-factor, the color hue is maintained all along while the chroma becomes higher, accompanied by the decreased net cooling power. It indicates that there is a trade-off between the net cooling power and the color chroma.

Based on the ideal CDRCs, a more realistic CDRC is considered by adopting a solar reflectance and mid-infrared emittance spectrum taken from the literature about the state-of-the-art DRC[27] and artificially introducing a reflectance dip at a wavelength of 435 nm for color display of yellow, as shown in Figure 2A. As reported, the DRC has an average solar reflectance of 0.86 at wavelengths of 0.3-4 μm and a mid-infrared emittance of 0.88 at wavelengths of 4-20 μm. Compared with the ideal CDRCs, the lower solar reflectance and mid-infrared emittance of the realistic CDRC would give rise to a more considerable absorbed solar irradiance. It is mandatory for the realistic CDRC to have a theoretical threshold of Q-factor higher than that of the ideal CDRCs. Following a similar procedure, the theoretical threshold of the realistic yellow CDRC can be determined to be 9.06 for sub-ambient cooling, as depicted in Figure 2B (see Supplementary Figure S4 for the realistic magenta and cyan CDRCs). According to the analysis, it can be understood that the MIM-based structure with a Q-factor of 5.72, extracted from the literature[18], cannot directedly cool to sub-ambient temperatures unless exploiting a spatial pattern, equivalent to the reduced colored area and absorbed solar irradiance.[28]

**Photonic design**

Central to achieve sub-ambient CDRCs is the Q-factor of reflectance dip in the visible light range exceeding the determined theoretical threshold while maintaining a high reflectance in the solar spectrum elsewhere and a high emittance in the mid-infrared range. Instead of the MIM or the absorbing dielectric/metal photonic structures, we consider TPRs for achieving the high-Q reflectance dip and maintaining high reflectance elsewhere. TPRs can be excited at the interface between a metal layer and a dielectric distributed Bragg reflector (DBR),[26] the Q-factor of which can be engineered by elaborately designing the TPR photonic structure.[29,30] More importantly, TPRs, unlike surface plasmon polariton resonances, can be directly excited on a homogenous surface without the need of sophisticated nanostructures for momentum matching, which is appealing to prospective scalable fabrication of CDRCs and practical applications. We verify these concepts by designing the TPR-based CDRCs to fulfill the stringent requirements of the high-Q reflectance dip in the visible light range, the high reflectance elsewhere in the solar spectrum, and the high emittance in the mid-infrared range. The TPR-based CDRCs have a polymer-Tamm photonic structure, which is composed of an organic thermal emission unit and an inorganic Tamm unit for achieving the desirable emittance and solar reflectance spectra, respectively, as depicted in Figure 3A.

To generate a reflectance dip with a Q-factor higher than the theoretical threshold for sub-ambient radiative cooling, we employ an aperiodic Tamm unit, in which the DBR does not follow the conventional quarter-wave stacking protocol for further enhancing the Q-factor of the reflectance dip[31] and a silver or aluminum metal layer has a thickness around 100 nm, much larger than the skin depth, for suppressing the optical losses and eliminating the transmission of solar irradiance (see Supplementary Figure S5 for the reflectance spectra of the periodic Tamm unit that adopts the conventional quarter-wave stacking protocol and a relatively thin metal layer for comparison). Herein, the DBR unit consists of four pairs of the alternating $Nb_2O_5$ and $SiO_2$ with negligible solar absorption, which have a large refractive index contrast, benefiting the generation of a broadband photonic bandgap and the strong confinement of TPRs at the interface between DBR and metal layer.[29] The reflectance dips are generated through exciting the TPRs, the spectral positions of which are tailored by the layer thickness of the dielectric distributed Bragg reflector, rendering the various structural colors. As an example, the magenta color is achieved when the thicknesses of four pairs of the alternating $Nb_2O_5$ and $SiO_2$ layers are set to be 35, 85, 63, 80, 60, 110, 50, and 20 nm from bottom to top as well as a 100-nm-thick Ag layer. Besides, the yellow and cyan colors are generated by using the different thickness configurations, as listed in Table S1. The theoretically calculated reflectance spectra are plotted in Figure 3B, where the reflectance dips appear at 435, 545, and 612 nm for the colors of yellow, magenta, and cyan, respectively. The designed colors in the chromaticity diagram are shown in Figure S6. As summarized in Table 1, the reflectance dips for generating the colors of yellow, magenta, and cyan have the Q-factors of 39.5, 38.9, and 21.1, respectively, all of which exceed the respective theoretical threshold of the Q-factor for sub-ambient radiative cooling. It reveals that the inorganic Tamm unit is eligible for color display at the lowest cost of the introduced thermal load.

Individually, the Tamm unit functions as a colored low-emissivity film that is highly reflective in the mid-infrared range and selectively reflective in the visible spectral range (see Figure S7), which is alternatively applicable for building walls for energy savings as colored low-emissivity films.[10] The Tamm unit, serving as a flexible platform, can be integrated with versatile organic polymers that are visibly transparent for elaborately manipulating the mid-infrared thermal emission, mechanical durability, and even surface wettability without affecting the high-Q TPRs and the color display. Organic polymers have specific functional groups, the vibrations of which are responsible for mid-infrared absorption/emission at specific wavelengths. We select polyvinyl alcohol (PVA) as a broadband thermal emitter, working in the mid-infrared spectral range at wavelengths of 5-20 μm due to the existence of functional groups of –OH (3225 cm$^{-1}$, i.e., 3.1 μm), C–OH (1825 to 1725 cm$^{-1}$, i.e., 5.5-5.7 μm), –CH (1382 to 1300 cm$^{-1}$, i.e., 7.2-7.7 μm , C–C (867.67 cm$^{-1}$, i.e., 11.5 μm) and –CH$_2$–(729 cm$^{-1}$, i.e., 13.9 μm) in PVA; for the selectively emissive CDRCs, PMMA can be adopted due to the existence of functional groups of C–O (1042 to 1267 cm$^{-1}$, i.e., 7.9-9.6 μm), and C–C (867.67 cm$^{-1}$, i.e., 11.5 μm) bonds in PMMA,[32,33] the bending vibrations wavelengths of which are mainly located within the atmosphere transparency window. According to the optical constants of the building blocks for the CDRCs (i.e., $Nb_2O_5$, $SiO_2$, Ag, Al, and PMMA) (see Supplementary Figure S8), the solar reflectance and infrared emittance spectra of the yellow CDRC with a broadband thermal emitter are calculated, as illustrated in Figure 3C (see Supplementary Figure S9 for those of the magenta and the cyan CDRCs). Due to the highly transparency of the organic unit in the solar spectral range from 0.3-2.5 μm, the integration of the Tamm unit with the organic unit barely affects the reflection characteristics originating from TPRs while a broadband or a selective thermal emission is achieved in the CDRCs. The spectral parameters of the CDRCs based on a polymer-Tamm photonic structure are summarized in Table 1, revealing the marginal solar absorption ($\alpha_{mean} < 13\%$) and significant thermal emission ($\varepsilon_{mean} \approx 0.95$).

**Fabrication and Optical Property Characterization of the CDRCs**

According to the theoretical analysis and photonic design above, the CDRCs were experimentally fabricated following the procedure illustrated in Figure 4A. The inorganic Tamm unit for both coloring and rejecting the solar irradiance was firstly fabricated by a magnetron sputtering deposition technique; the PVA or PMMA polymer unit was uniformly coated onto the Tamm unit by a chemically spin-coating method, followed by thermal curing of the polymer unit at a temperature of 45 °C, not only achieving broadband or a selective thermal emission in the infrared spectral range but also protecting the CDRCs from external mechanical impact as an encapsulation. The fabrication is completely scalable through the magnetron sputtering deposition and the chemical coating techniques, which can be realized by industrial R2R facilities.

The measured reflectance spectra of the Tamm unit in the visible spectral range were characterized and compared with the theoretical designs, as illustrated in Figure 4B-C. In terms of the central wavelength of the reflection dip $\lambda_{max}$ and the corresponding Q-factor, the measured results are in accordance with the calculated. The central wavelengths of the yellow, magenta, and cyan Tamm unit are located at 447, 550, and 623 nm, corresponding to the Q-factor values of 23.5, 34.3, and 20.7, respectively, all of which surpass the respective theoretical threshold for sub-ambient cooling. Figure 4D demonstrates the photographs of the samples based on the Tamm unit, which reveals the vivid subtractive primary colors of yellow, magenta, and cyan. The cross-sectional SEM of the magenta sample (Figure 4E) reveals that the thicknesses of the magenta Tamm unit, consisting of four pairs of the alternating $Nb_2O_5$ and $SiO_2$ layers, are 45, 79, 60, 77, 62, 114, 57, and 29 nm from bottom to top as well as a 102-nm-thick Ag layer, corresponding to a relatively small deviation of 5 nm on average from the theoretical design.

These samples based on the individual Tamm unit without the polymer unit are initially low-emissive in the infrared range, as characterized in Supplementary Figure S10. The Tamm unit, as a platform, can be integrated with versatile organic polymers that are visibly transparent and mid-infrared selectively or broadband emissive for elaborately manipulating thermal radiation, mechanical durability, and so forth without affecting the color display. Thus, the PVA layer with a thickness around 15 μm was coated onto the Tamm unit, which is denoted as CDRC-PVA for clarity. Figure 4F presents the measured reflectance spectra in the solar spectral range from 0.3-2.5 μm for the yellow, magenta, and cyan CDRC-PVAs. It reveals that the involvement of the PVA polymer unit has few effects on the TPRs, since the TPRs are independent of the polymer layer and locally confined at the interface between the $Nb_2O_5/SiO_2$ pairs and the metal layer instead. Meanwhile, a high reflectance is still preserved in the solar spectrum elsewhere due to the high transparency of PVA and the effects of the PVA on the reflectance spectrum of the Tamm unit are marginal (see Supplementary Figure S11). Figure 4G presents the measured absorptance/emittance spectra in the infrared spectral range from 2.5-20 μm. Due to the involvement of the PVA polymer unit, the average emittances in the atmospheric transparency window from 8 to 13 μm are increased from 0.07, 0.12, and 0.30 in the case of individual Tamm unit to 0.91, 0.94, and 0.92 for the yellow, magenta, and cyan CDRC-PVAs, respectively. Alternatively, the PMMA polymer unit can also be integrated with the Tamm unit, which is named CDRC-PMMA (see Figure S12, Supporting Information).

**Field Measurements of Radiative Cooling Performance**

Figure 5A (left) shows the photographs of the setup, together with the CDRC samples, for the field 24-h continuous temperature measurements during a winter day and night, which was conducted on the rooftop of a building in Ningbo city (28.51 °N, 120.55 °E), China, in November-December 2021. The measurement setup was supported by an acrylic box for reducing heat conduction from the ground, as schematically illustrated in Figure 5A (right). The measurement chamber, made from 3-mm-thick transparent acrylic plates, had a size of 30 cm × 30 cm × 8 cm in length × width × height, covered by the aluminum foils for rejecting the incident solar irradiance and an infrared-transparent low-density polyethylene film on the top for radiating heat out and preventing convection. The CDRC samples were placed on the polystyrene mats that were also wrapped by the aluminum foils. The thermocouples were attached to the back of the samples for determining the sample's temperature ($T_{sample}$) while the other thermocouple was placed at the same height to measure the inner ambient temperature ($T_{ambient}$). The temperature data were logged at every 10s by a data logger (RDXL4SD, Omega Engineering, USA). The outer ambient temperature and relative humidity were recorded every 30s by a temperature/humidity sensor (SHT 31, Sensirion,

Switzerland) and the solar irradiance was logged every 60s by a weather station (PC-4, Jinzhou Meteorological, China).

The results of outdoor measurements over 24h on Nov. 25-26, 2021, for the CDRC-PVAs are shown in Figure 5B-E. The measured temperatures of the inner ambient, the yellow, magenta, and cyan CDRC-PVAs, together with the solar irradiance, are plotted in Figure 5B. Figure 5C exhibits the outer temperature and the relative humidity during the measurements: the outer temperature varies from 15 °C to 30 °C, and the relative humidity varies from 18% to 90% during the daytime and nighttime, both of which change in an opposite way. Comparing the inner and outer ambient temperatures, the former was higher than the latter during the daytime and the latter becomes higher during the nighttime due to the greenhouse effect inside the measurement chamber[34] and the suppressed heat convention. The variations of the inner ambient and the sample temperatures basically follow the solar irradiance intensities. During the daytime of the first day between 10:00 and 16:00 local time, the average temperatures of the yellow, magenta, and cyan CDRCs were 21.1, 23.4, and 27.3 °C, lower than the average ambient temperature 29.9 °C when the solar irradiance was in the range of 100-735 W/m². And the temperatures of the three CDRC-PVAs were lower than the ambient temperature all along: the temperature differences of the yellow, magenta, and cyan CDRC-PVAs reached $\Delta T = T_{ambient} - T_{sample} = 8.8, 5.3$, and 1.6 °C, respectively, even under a peak solar irradiance of 735 W/m² at 10:58 local time on the first day, strongly confirming that sub-ambient radiative cooling is possible for CDRCs through sophisticated tailoring of the high-Q reflection dip for coloring in the visible range and thermal emission in the infrared range as well as maintenance of high reflectance elsewhere. It is noted that the yellow CDRC-PVA shows the most outstanding cooling performance with a maximum temperature difference $\Delta T = 11$°C under the solar irradiance of 612 W/m² and the cyan CDRC-PVA has the smallest temperature difference because it has a relatively low Q-factor of 20.76, below that of the yellow CDRC-PVA, 23.52. During the nighttime between 18:00 and 04:00, the temperature differences ΔT of the yellow, magenta, and cyan CDRC-PVAs were 4.4, 4.3, and 4.0 °C on average, which are comparable with the state-of-the-art non-colored radiative coolers in terms of nighttime radiative cooling performance and can be ascribed to their comparable infrared emittances. The measured average temperatures of the three CDRCs relative to the ambient air temperature in daytime and nighttime is shown in Figure 5D. It is worthwhile to note that the sub-ambient radiative cooling performance of the yellow and magenta CDRC-PVAs at night was (~4.3 °C) inferior to those at daytime (~7.8 °C, ~6.2 °C), because of the declining transparency of the atmospheric window as the increases of relative humidity at night. For the cyan CDRC-PVA, the sub-ambient radiative cooling performance at night was better than at daytime, which is understood that the adverse effects of the relatively large solar absorption due to the low-Q and the accompanied reflection dips on radiative cooling during the daytime are more significant than those of the declining transparency of the atmospheric window during the nighttime. The net cooling powers for the CDRCs were calculated using their respective solar reflectance and infrared emittance values in Figure 5E for determining the steady-state temperatures. For the yellow, magenta, and cyan CDRC-PVAs, the calculated steady-state temperatures are 19.8, 20.9, and 24.4 °C when taking $h_c$ = 6 W/m²/K , $T_{amb}$ = 30 °C and the solar irradiance is 700 W/m²; the measured results are 21.3, 24.3, and 28.7 °C, respectively, at an inner ambient temperature of 30 °C. The calculated steady-state temperatures are lower than the measured, which might be associated with the higher atmospheric transmittance used in the calculations than that in the field tests. Both the theoretical and experimental results strongly confirm that the CDRC has a good sub-ambient cooling capability. In addition, the cooling performance of CDRC-PMMAs with a selective thermal emitter (see Figure S13), also exhibiting the sub-ambient cooling ability. The cooling performance of CDRCs with either a broadband or a selective emitter was compared through calculating their cooling power values when taking $h_c$ = 6 W/m²/K , $T_{amb}$ = 30 °C and a solar irradiance of 1000 W/m² (see Figure S14).

**Scalable manufacture of CDRCs**

Besides the realization of vivid colors and sub-ambient radiative cooling, the scalable production of CDRCs remains highly challenging. We demonstrate the proposed CDRCs based on the polymer-Tamm photonic structure can be produced by the industrial R2R facilities. Figure 6A exhibits a 60-cm-wide and 500-cm-long CDRC, where the Tamm and the polymer units were prepared by a R2R magnetron sputtering deposition and blade coating, respectively. The anti-scratch performance of the CDRC was evaluated by an abrasion tester, as illustrated in Figure 6B, where the samples were rubbed by the felt under 200 g load. Comparing the appearances of the sample before and after being scratched for 2000 cycles, the magenta CDRC still maintained the similar color and no obvious scratches are observed (Figure 6C). The adhesion strength of the prepared polymer-Tamm photonic structure in the magenta CDRC was evaluated by a standard cross-cut tape test, according to ASTM D3359. As shown in Figure 6D, the edges of the cuts are completely smooth without any detached squares of the lattice. The adhesion strength can be classified as the highest level, 5B, implying that there is a strong adhesion between not only the photonic structure and the PET substrate, but also the polymer and the Tamm units. We further inspected the tensile properties of the magenta CDRC sample. The superior tensile properties are observed from Figure 6E, where the sample can withstand a strain of 152% and a maximum tensile force of 141 MPa, equivalent to 244 N for the tested sample with a size of 12.72 mm× 0.136 mm in width × thickness. The abrasion, adhesion, and tensile tests all validate the excellent mechanical stability of the CDRC sample, together with the vivid colors and sub-ambient radiative cooling performance, which would make the CDRCs be closer to reality.

In summary, we realize the yellow, magenta, and cyan CDRCs for all-day sub-ambient cooling, which are composed of a high-Q Tamm photonic structure for coloring at the lowest cost of absorbed solar irradiance and an organic transparent coating for working as a thermal emitter. The theoretical thresholds of Q-factor for the yellow, magenta, and cyan CDRCs with a broadband thermal emitter are determined to 9.06, 10.07, and 10.86, respectively. According to the theoretical guidance, we experimentally fabricate the CDRCs that not only exhibit the subtractive primary colors of yellow, magenta, and cyan, but also cool to 2.6-8.8 °C and 4.0-4.4 °C below the air temperature on average during daytime and nighttime, respectively. We further demonstrate the large-scale manufacturing of a 60-cm-wide and 500-cm-long magenta CDRC by using an R2R magnetron sputtering deposition and a blade coating. We believe that the hindrance removals of diverse colors, sub-ambient cooling, and scalable manufacture to CDRCs indicate unprecedented opportunities for elaborately engineering the absorbed solar irradiance with desirable colors and potential applications in building envelopes and vehicles that have energy-saving, aesthetic, and visual comfort demands.

**EXPERIMENTAL PROCEDURES**

**Resource Availability**

*Lead Contact*

Further information and requests for resources and reagents should be directed to and will be fulfilled by the Lead Contact, Yuehui Lu (yhlu@nimte.ac.cn).

*Materials Availability*

This study did not generate new unique materials.

*Data and Code Availability*

All of the data associated with the study are included in the article and the Supplemental Information. Additional information is available from the Lead Contact upon reasonable request.

**Materials**

Polyethylene terephthalate was bought from Sinopharm Chemical Reagent Co, Ltd. Polyvinyl alcohol (P105126), Polymethyl methacrylate (P141442) were purchased from Sigma-Aldrich, and trichloromethane (99.9%) were purchased from Sinopharm Chemical Reagent Co, Ltd. Ag (99.999%), $SiO_2$ (99.99%), and $Nb_2O_5$ (99.99%) targets with a diameter of 3 in. were used as sputtering targets.

**Fabrication of Tamm unit.**

Tamm unit was fabricated using large-scale R2R sputtering system on 600-mm-wide commercial polyethylene terephthalate (PET) substrate. The Ag layer was deposited using an Ag target with DC power of 2 KW under 400 sccm Ar flow rate. The $Nb_2O_5$ layers were sputtered using a $Nb_2O_5$ target at a DC power of 3 KW, $Ar/O_2$ flow ratio of 400/30 sccm. The $SiO_2$ layers were reactive sputtered using a Si target at a mid-frequency power of 6 KW, $Ar/O_2$ flow ratio of 400/100 sccm. The thickness of each layer was controlled through the rolling speed of the PET substrate.

**Fabrication of PVA layer**

The PVA powder and deionized water were mixed with a mass ratio of 13:87 and placed in an oil bath at 90 °C for 1 h until the PVA powder dissolved completely. The air bubbles are removed during the ultrasonication process. The uniform PVA layer was deposited on the Tamm film by means of spin coating technique at speed of 400 rcf. The Tamm film was then dried for 12 h at 40°C.

**Characterization of Optical Properties.**

The solar reflectance from 0.2um to 2.5 um was measured with an ultraviolet−visible−near-IR (UV−Vis−NIR) spectrometer (Perkin-Elmer, Lambda 1050) with an integrating sphere. The reflectance ($R$) and transmittance ($T$) spectra from 2.5 to 20 μm were measured using a Fourier transform infrared (FTIR) spectrometer (Nicolet 6700, Thermo, USA) with an integrating sphere and the absorptivity/emissivity value was obtained using the equation of $1 - R - T$.

**Principles of radiative cooling**

The net cooling power is expressed by

$$P_{\text{cool}} = P_{\text{rad}}(T) - P_{\text{atm}}(T_{\text{amb}}) - P_{\text{sun}}(\theta) - P_{\text{cond+conv}} \quad (1)$$

where $T$ is the surface temperature of the radiative cooler and $T_{\text{amb}}$ is the ambient temperature. $P_{\text{rad}}(T)$ is the radiative power of the cooler, and $P_{\text{atm}}(T_{\text{amb}})$ is the absorbed atmospheric thermal radiation at $T_{\text{amb}}$. $P_{\text{sun}}(\theta)$ is the power of solar irradiation absorbed by the cooler, and $P_{\text{cond+conv}}$ is the power of nonradiative heat exchange due to convection and conduction. These parameters can be calculated by the following equations:

$$P_{\text{rad}}(T) = \int d\Omega \int_0^\infty I_{\text{BB}}(T,\lambda)\varepsilon(\lambda,\theta)\cos\theta\, d\lambda \quad (2)$$

where $\int d\Omega = \int_0^{\frac{\pi}{2}} d\theta \sin\theta \int_0^{\frac{\pi}{2}} d\phi$ is the angular integral over a hemisphere, $I_{\text{BB}}(T,\lambda) = 2hc^2/\lambda^5/[e^{hc/(\lambda k_{\text{B}} T)} - 1]$ is the spectral radiance of a blackbody at temperature T and $h$, $c$, and $k_{\text{B}}$ are the Planck's constant, the light speed, and the Boltzmann constant, respectively. $\varepsilon(\lambda,\theta)$ is the wavelength- and angle-dependent emissivity of the radiative cooler.

$$P_{\text{atm}}(T_{\text{amb}}) = \int d\Omega \int_0^\infty I_{\text{BB}}(T_{\text{amb}},\lambda)\varepsilon(\lambda,\theta)\,\varepsilon_{\text{atm}}(\lambda,\theta)\cos\theta\, d\lambda \quad (3)$$

where $\varepsilon_{\text{atm}}(\lambda, \theta) = 1 - t(\lambda)^{1/\cos\theta}$ is the emissivity of the atmosphere and $t(\lambda)$ is the atmosphere transmittance in the zenith direction.

$$P_{\text{sun}}(\theta) = \int_0^\infty \varepsilon(\lambda, \theta) I_{\text{AM1.5}}(\lambda) d\lambda \tag{4}$$

where $I_{\text{AM1.5}}(\lambda)$ represents the AM1.5 solar illumination with a total solar irradiance of about 1000 W/m².

$$P_{\text{cond+conv}} = h_c(T_{\text{amb}} - T) \tag{2}$$

where $h_c$ is the combined nonradiative heat coefficient of convection and conduction.

**Principles of color display**

We use the CIE 1931 color coordinate system based on the *CIE-XYZ* coordinate space to predict the color of the Tamm film via the simulated reflectance system. The color information was expressed by X, Y, and Z, which are the tristimulus values. The expressions of the tristimulus values of the object color are

$$X = \int R(\lambda)\bar{x}(\lambda)\, d\lambda$$

$$Y = \int R(\lambda)\bar{y}(\lambda)\, d\lambda$$

$$Z = \int R(\lambda)\bar{z}(\lambda)\, d\lambda$$

where the $R(\lambda)$ is the reflectance spectrum of the object, $\bar{x}(\lambda), \bar{y}(\lambda), \text{and}\, \bar{z}(\lambda)$ are the color matching functions, which numerically represented the chromatic sensitivities of the human eyes. Then the color chromaticity is determined by the normalized parameters $x, y$, and $z$:

$$x = \frac{X}{X + Y + Z}$$

$$y = \frac{Y}{X + Y + Z}$$

$$z = \frac{Z}{X+Y+Z}$$

The color is determined by the chromaticity coordinate (*x*, *y*) in the chromaticity diagram, which, in turn, is determined by its reflection spectrum in the 1931 *CIE-XYZ* color system. To better describe the chromatic response of human eye, the *CIE-L*a*b**, color space is put forward by the nonlinear transformation of *CIE-XYZ* color space. Here $L^*$ represents the lightness, $a^*$ represents the redness and greenness, and $b^*$ represents the yellowness and blueness. These parameters can be transformed from the three tristimulus value X, Y, Z numerically as:

$$L^* = 116 f(Y/Y_0) - 16$$

$$a^* = 500[f(X/X_0) - f(Y/Y_0)]$$

$$b^* = 200[f(Y/Y_0) - f(Z/Z_0)]$$

where $X_0, Y_0$, and $Z_0$ are the tristimulus values corresponding to color white, and

$$f(t) = \begin{cases} t^{1/3}, & t > (\frac{24}{116})^3 \\ \left(\frac{841}{108}\right)t + \frac{16}{116}, & t \leq (\frac{24}{116})^3 \end{cases}$$

We use the *CIE-LCH* color space to illustrate color quality more intuitively. It uses lightness ($L = L^*$), chroma ($C = \sqrt{a^{*2} + b^{*2}}$), and hue ($H = \arctan(b^*/a^*)$) to describe colors, which fits better with human visual experience.

**SUPPLEMENTAL INFORMATION**

Supplemental Information can be found online at https://doi.org/.

**ACKNOWLEDGMENTS**

This work was supported by the National Natural Science Foundation of China (61875209 and 12174209), the Natural Science Foundation of Zhejiang Province (LY19F040003), the Ningbo Key Laboratory of Silicon and Organic Thin Film Optoelectronic Technologies, and the Jiangsu Province Cultivation base for State Key Laboratory of Photovoltaic Science and Technology.

**AUTHOR CONTRIBUTIONS**

**DECLARATION OF INTERESTS**

The authors declare no competing interests.

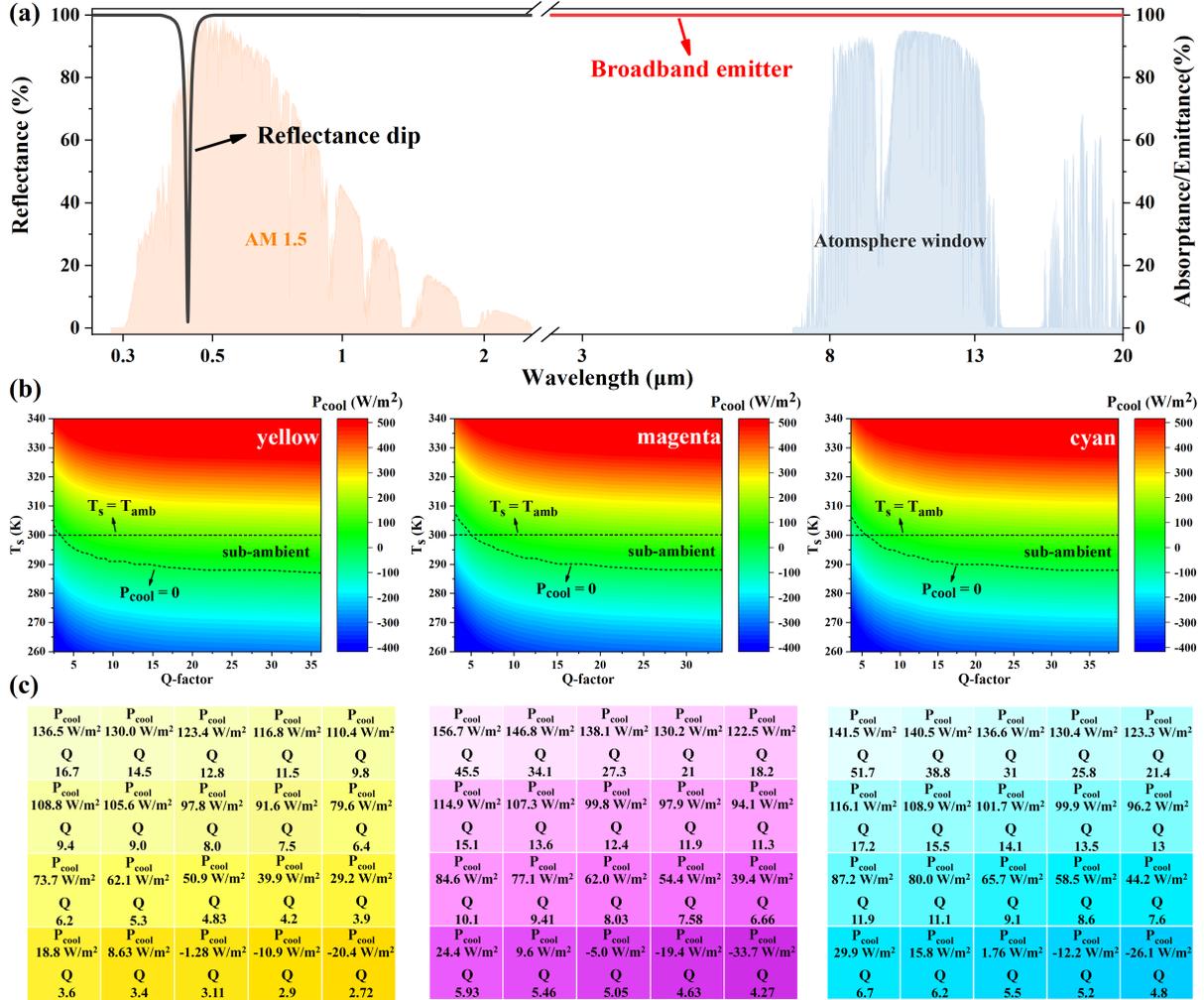

**Figure 1. Concept illustration of ideal CDRCs for sub-ambient cooling**

(A) Reflectance (black line) and absorptance/emittance spectra of a representative ideal yellow CDRC with a broadband thermal emission. The normalized AM 1.5G solar and the atmospheric transmittance spectra are illustrated for reference.

(B) Net cooling power of yellow, magenta, and cyan CDRCs with a broadband thermal emission as a function of Q-factors of the reflectance dips and their surface temperatures for a realistic condition of $h_c$ = 6 W/m²/K and $T_{amb}$ = 300 K.

(C) Their dependences of chroma and hue on the Q-factors and the corresponding net cooling power.

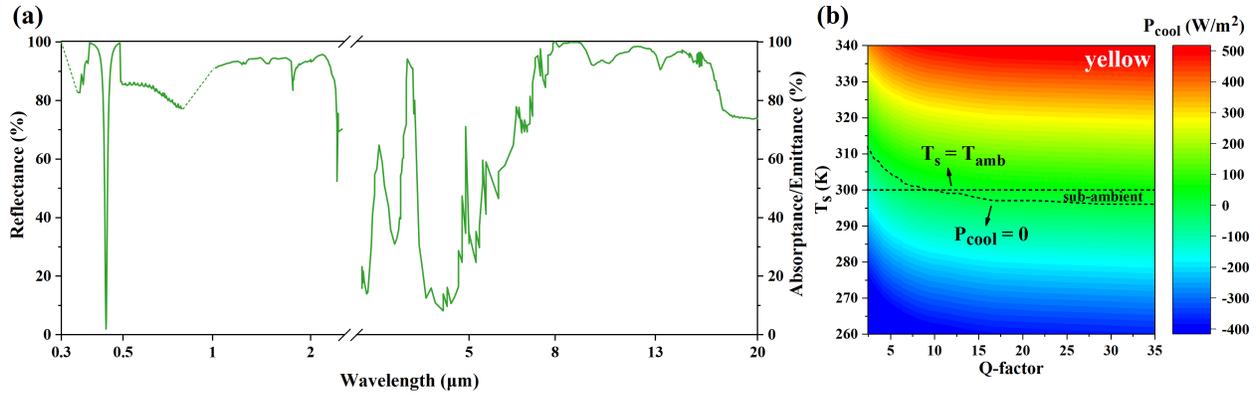

**Figure 2**. **Theoretical analysis of a CDRC modified from a state-of-the-art DRC**

(A) Introducing a reflectance dip into the state-of-the-art DRC for yellow color display.

(B) Net cooling power of the CDRC as a function of Q-factors of the reflectance dips and their surface temperatures for a realistic condition of $h_c$ = 6 W/m²/K and $T_{amb}$ = 300 K.

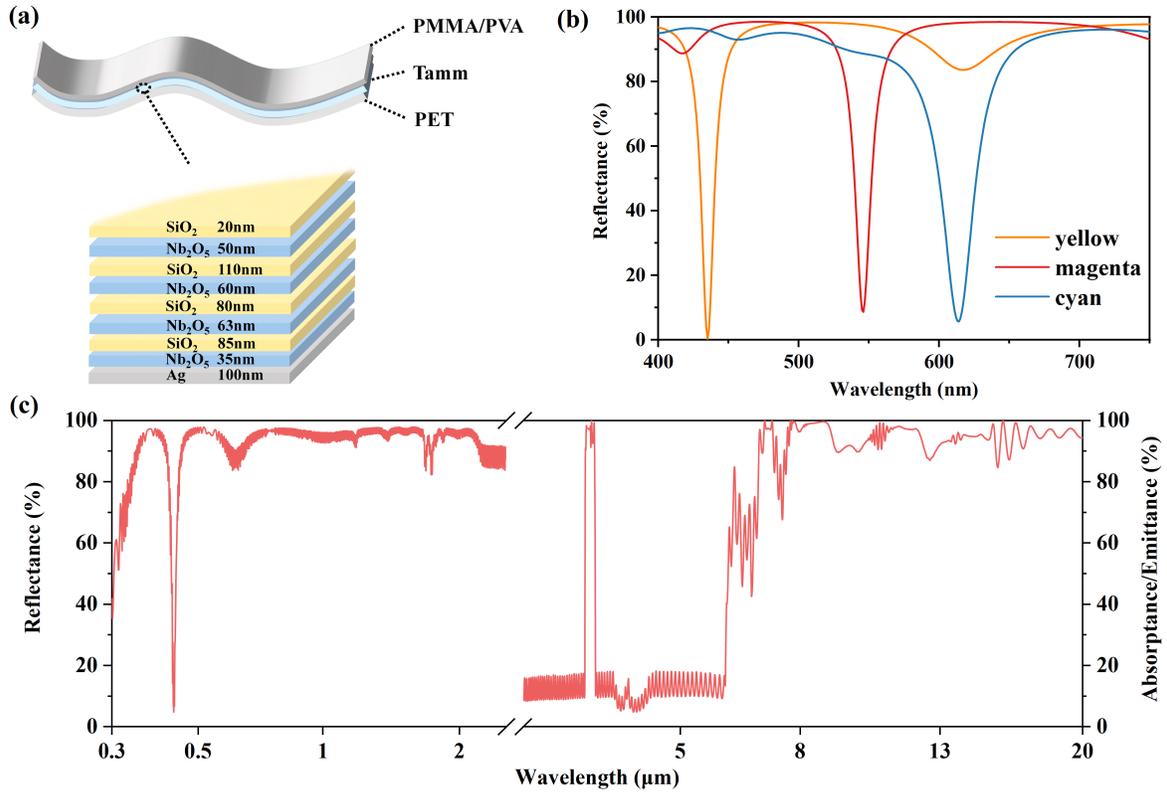

**Figure 3. Theoretical design of CDRCs based on a polymer-Tamm photonic structure**

(A) Schematic illustration of the CDRCs, composed of an aperiodic Tamm structure with high Q-factors and an organic multifunctional layer for the purpose of tailoring the thermal emissivity and enhancing the durability.

(B) Calculated reflectance spectra of the designed aperiodic Tamm structures in yellow, magenta, and cyan in the visible region.

(C) Calculated emissivity spectra of the yellow CDRC with the broadband thermal emission.

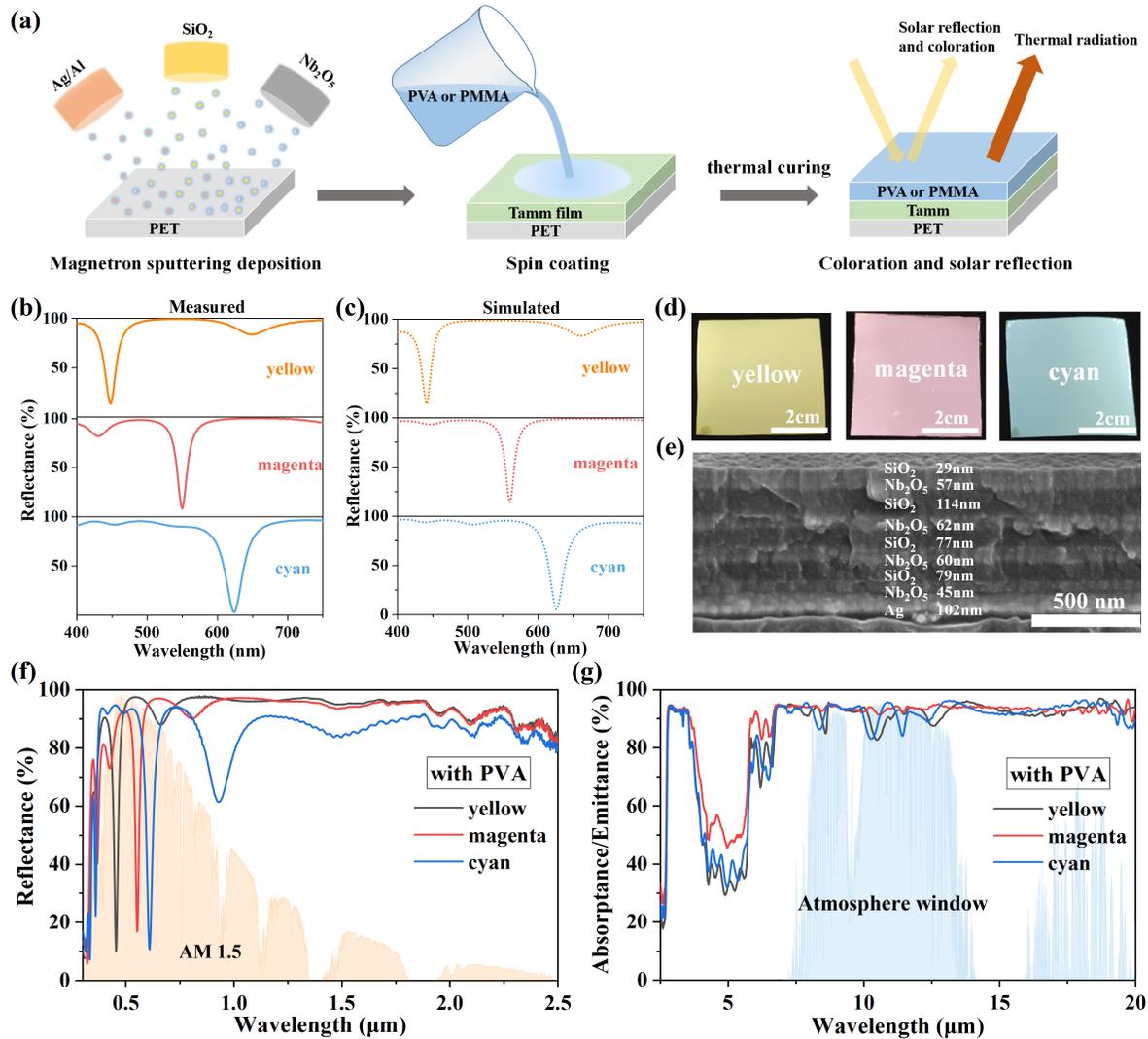

**Figure 4. Fabrication and characterization of the CDRCs**

(A) Schematic illustration of the experimental realization of the CDRCs.

(B and C) Measured (B) and simulated (C) reflectance spectra of the aperiodic Tamm units in yellow, magenta, and cyan.

(D) Photographs of the three fabricated aperiodic Tamm units. Scale bar, 2 cm.

(E) Cross-sectional SEM image of the magenta Tamm unit. Scale bar, 500 nm.

(F and G) Measured solar reflectance spectra (F) and IR emissivity spectra (G) of the CDRCs with the broadband PVA thermal emitter in yellow, magenta, and cyan.

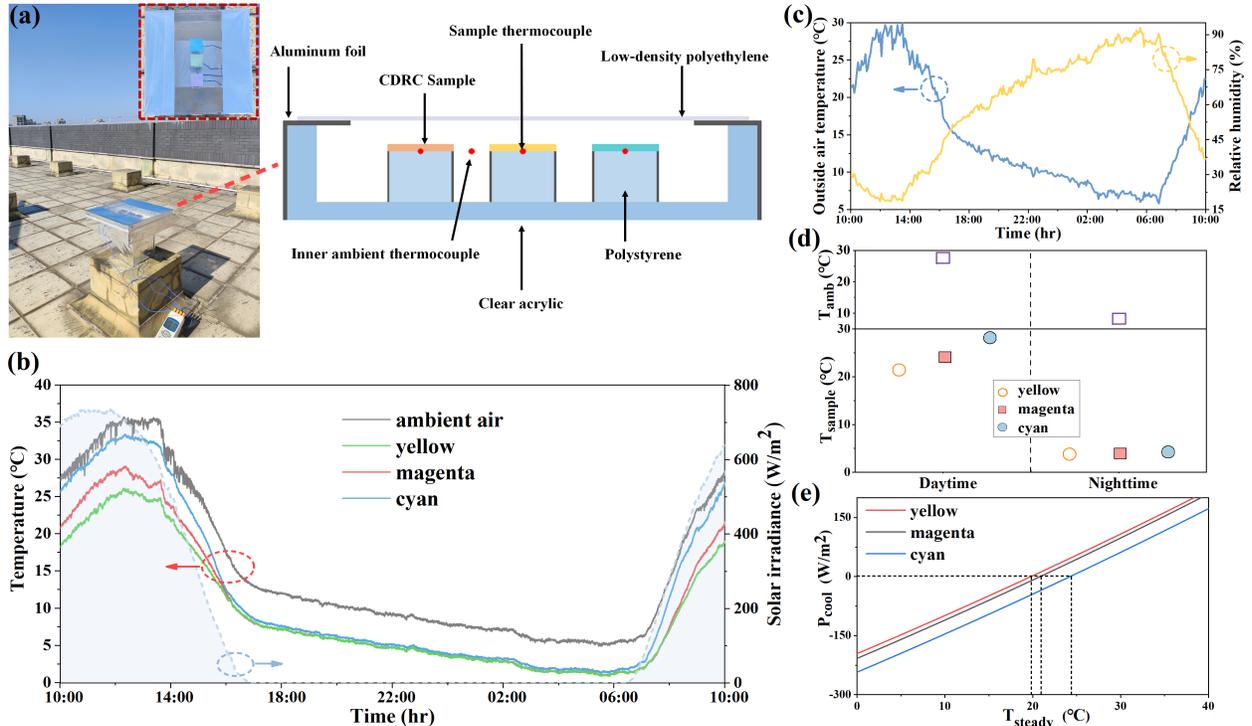

**Figure 5. Radiative cooling performance of the CDRC-PVAs.**

(A) Photographs of the setup for measuring the radiative cooling performance of the CDRC samples during the daytime and nighttime. Inset: photograph of three CDRC samples in yellow, magenta, and cyan under sunlight. Schematic illustration of the setup for the field temperature measurement.

(B) Measured temperatures of the broadband emissive yellow, magenta, and cyan CDRCs over 24 h on Nov. 25—26, 2021, together with the solar irradiance (blue area).

(C) Measured relative humidity and outside ambient temperature.

(D) Measured average temperatures of the broadband emissive CDRCs relative to the ambient air temperature in daytime and nighttime.

(E) Calculated cooling power of the yellow, magenta, and cyan CDRCs.

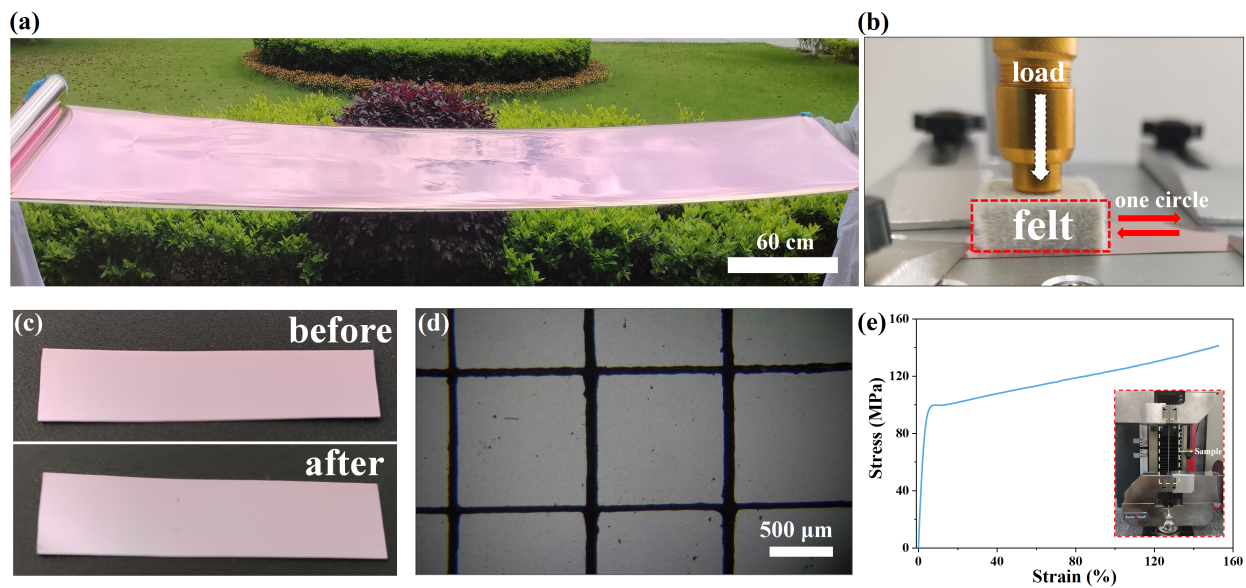

**Figure 6. Scalable manufacture and mechanical characterization of the magenta CDRC**

(A) Photograph of the large-scale magenta CDRC sample with a size of 60 cm width by 500 cm length.

(B) Photograph of the linear abrasion tester. Samples are rubbed by the felt with a certain weight load. One circle represents a forward and backward movement with a distance of 10 cm.

(C) Photographs of the CDRC sample before and after an abrasion test by rubbing the sample surface with a felt under 1.96 N of the pressure.

(D) SEM image of the CDRC sample after a standard cross-cut tape test.

(E) Mechanical tensile test for the CDRC sample.

**Table 1. Summarized spectral parameters of the designed yellow, magenta, and cyan CDRCs.**

| Spectral parameters | Yellow | Magenta | Cyan |
|---|---|---|---|
| $\lambda_{max}$ | 435 nm | 545 nm | 612 nm |
| $\Delta\lambda$ | 11 nm | 14 nm | 29 nm |
| Q-factor | 39.5 | 38.9 | 21.1 |
| $\alpha_{mean}$ (0.3-2.5μm) | 0.06 | 0.08 | 0.13 |
| $\varepsilon_{mean}$ (8-13μm) | 0.95 | 0.95 | 0.95 |

**Table 2. Summarized average absorptances of the CDRCs in the solar and the LWIR atmospheric window spectral range.**

| Color | Yellow | | Magenta | | Cyan | |
|---|---|---|---|---|---|---|
| $\alpha_{mean}$ (0.3-2.5μm) | 0.08 | 0.09 | 0.09 | 0.09 | 0.16 | 0.16 |
| $\varepsilon_{mean}$ (8-13μm) | 0.91 | 0.81 | 0.93 | 0.81 | 0.92 | 0.86 |